\documentclass[12pt, reqno]{amsart}
\usepackage{graphicx}
\usepackage{subcaption}
\usepackage{amsfonts,amsmath,amsthm,amssymb,color}
\usepackage{mathtools}
\usepackage{fullpage}
\usepackage{hyperref}

\usepackage[backend=biber,sorting=nty,style=numeric,maxbibnames=99]{biblatex}
\title{Modeling Processes of Neighborhood Change}
\date{\today}

\author[Mart\'inez Mori]{J. Carlos Mart\'inez Mori}
\address[J.~C. Mart\'inez Mori]{H. Milton Stewart School of Industrial and Systems Engineering, Georgia Institute of Technology}
\email{\textcolor{blue}{\href{mailto:jcmm@gatech.edu}{jcmm@gatech.edu}}}

\author[Zhao]{Zhanzhan Zhao}
\address[Z. Zhao]{Simons Laufer Mathematical Sciences Institute}
\email{\textcolor{blue}{\href{mailto:zhanzhanzhao@gatech.edu}{zhanzhanzhao@gatech.edu}}}

\begin{document}


\begin{abstract}
An urban planner might design the spatial layout of transportation amenities so as to improve accessibility for underserved communities\textemdash a fairness objective.
However, implementing such a design might trigger processes of neighborhood change that change who benefits from these amenities in the long term.
If so, has the planner really achieved their fairness objective?
Can algorithmic decision-making anticipate second order effects?
In this paper, we take a step in this direction by formulating processes of neighborhood change as instances of no-regret dynamics; a collective learning process in which a set of strategic agents rapidly reach a state of approximate equilibrium.
We mathematize concepts of neighborhood change to model the incentive structures impacting individual dwelling-site decision-making.
Our model accounts for affordability, access to relevant transit amenities, community ties, and site upkeep.
We showcase our model with computational experiments that provide semi-quantitative insights on the spatial economics of neighborhood change, particularly on the influence of residential zoning policy and the placement of transit amenities.
\end{abstract}

\maketitle

\section{Introduction}

The design of transportation infrastructure is studied from a wide variety of disciplines, from applied mathematics, engineering, and computer science to city and regional planning.
For example, in technical disciplines, one might abstract the problem of placing the stations of a proposed subway line as a kind of \emph{facility location problem}.
Recently, there has been growing interest in incorporating notions of ``fairness'' into these abstractions\textemdash a task at the intersection of planning practice, theories of justice, and mathematical optimization~\cite{xinying2023guide}.
The facility location literature (broadly construed) showcases various recent examples~\cite{filippi2021single,gupta2023lp,abbasi2021fair,banerjee2020optimal,hickok2022persistent}.
To formalize these decisions, one generally takes some pre-specified, snapshot form of input (e.g., the spatial demography of potential transit users), from which an ``optimal'' solution is subsequently derived and implemented.
However, by definition, infrastructure planning has real, large scale impact on real people.
In particular, the \emph{status-quo can change in response to transportation infrastructure}, possibly rendering assumptions on the problem input inadequate and casting doubt on the ``optimality'' of derived solutions.

This paper is motivated by the following problem.
An urban planner might design (e.g., using mathematics-enabled software) the spatial layout of transportation amenities so as to improve accessibility for underserved communities\textemdash a ``fairness'' objective.
However, implementing such a design might trigger processes of \emph{neighborhood change} that change \emph{who} benefits from these amenities in the long term.
If this holds, has the planner really achieved their ``fairness'' objective?
Can algorithmic decision-making anticipate second order effects?
Here, we take a first step in this direction by applying techniques from algorithmic game theory within the context of transportation and processes of neighborhood change.

\subsection{Neighborhood Change}
\label{sec: neighborhood change}

The term \emph{gentrification} was coined by Glass~\cite{glass1964london} to describe the movement of the ``gentry'' into working-class neighborhoods of London during the mid-twentieth century.
Since then, social scientists have strived to develop a general theory of neighborhood change.
This is a formidable goal: comprehensive definitions are elusive and, partly in consequence, supporting data is largely unavailable. 

Gentrification generally refers to a \emph{process} of neighborhood renewal in which the primary beneficiaries tend to be homeowners and newcomers, as opposed to incumbent renters~\cite{zuk2018gentrification}.
It is therefore not a static binary attribute (i.e., is this neighborhood gentrified?).
Instead, the concept necessitates a longitudinal perspective.
Similarly, it does not carry an inherent value or operative judgment.
That is, depending on the context and standpoint, gentrification may be deemed positive or negative; as a tool, consequence, or goal~\cite{lees2008gentrification,kohn2013wrong,zuk2018gentrification}.

However, it is generally understood that gentrification can result in \emph{displacement}.
Grier and Grier~\cite{grier1978urban} define residential displacement as the the forced movement of a household because of circumstances affecting their dwelling or its surroundings, including but not limited to official action (e.g., an eviction).
More broadly, displacement arises from any external stressors that negatively impact a household such as dwelling and/or neighborhood deterioration (e.g., as a result of redlining), rent and/or property tax hikes, and loss of a sense of community/social ties~\cite{newman1982residential,marcuse1985gentrification, davidson2009displacement}.
These forms of displacement can but need not lead to spatial dislocation.
For example, as concluded by Freeman and Braconi~\cite{freeman2004gentrification}, ``gentrification brings with it neighborhood improvements that are valued by disadvantaged households, and they consequently make greater efforts to remain in their dwelling units, even if the proportion of their income devoted to rent rises.''

Gentrification and displacement are distinct but intertwined concepts.
Displacement may precede gentrification in the form of dwelling and/or neighborhood deterioration, just as it may accompany it in the form of social and/or financial stress.
Indeed, gentrification is often thought to arise in waves~\cite{hackworth2001changing,lees2008gentrification,owens2012neighborhoods}, each with different incentives and socioeconomic implications.
These range from the \emph{urbanization of poverty}~\cite{glaeser2008poor}, characterized by the decline of the urban core and car-centric white flight, to the \emph{suburbanization of poverty}~\cite{sheller2015racialized,kramer2018unaffordable}, characterized by the renewal of the urban core and increased transportation burdens for marginalized populations.
Hackworth and Smith~\cite{hackworth2001changing} argue that the first wave of newcomers into a gentrifying neighborhood consists of highly-educated individuals who are driven by its (relative) affordability, amenities, and culture, whereas subsequent waves are characterized by the increasing presence of large-capital, often with implicit government support.
The latter may be in the form of direct investment, \emph{notably transportation infrastructure}, as well as in the form of indirect policy, notably zoning regulations~\cite{zuk2018gentrification,kramer2018unaffordable,chapple2019transit}.
In the United States, the racial wealth-gap and structural patterns of racism and segregation compound the conceptual nuance with a racial connotation~\cite{sheller2015racialized,fallon2021reproducing,rucks-ahidiana2022theorizing}.

\subsection{Contributions}
\label{sec: contributions}

The complexity and inconspicuousness of processes of neighborhood change make them and their interactions with policy interventions extremely difficult to quantify~\cite{atkinson2000measuring,newman2006right}.
In this paper, we propose the use of game-theoretic concepts to grapple with some of these technical difficulties.
Our contributions are as follows:
\begin{enumerate}
    \item 
    We formulate processes of neighborhood change as instances of \emph{no-regret dynamics}, wherein a set of strategic agents provably converge to a state of approximate equilibrium by way of repeated play.
    Compared to existing agent-based models in the study of urban affairs, this approach treats agents as strategic players instead of assuming they behave as mechanical particles.
    
    In our model, we consider a set of economically diverse residents competing for a set of spatially diverse dwelling sites.
    We mathematize concepts outlined in Section~\ref{sec: neighborhood change} to capture the incentive structures perceived by any given resident as a function of the dwelling-site choices of all other residents.
    Our model explicitly accounts for housing affordability, access to transit amenities, community ties (with respect to a class-based conception), and dwelling site abandonment/upkeep.  
    \item 
    We showcase our model with computational experiments that shed some light on the spatial economics of transit-related neighborhood change.
    We use publicly available tools~\cite{boeing2017osmnx} to obtain a network representation of a real-world urbanized region with a non-trivial configuration of transit stations.
    We study the effects of residential zoning policy and resident preferences over transportation amenity access and community ties.
    Our experimental results suggest the following:
    \begin{enumerate}
        \item 
        the combination of transit amenities and low-density zoning results in the suburbanization of poverty,
        \item
        the combination of transit amenities and medium-density zoning results in financially heterogeneous neighborhoods concentrated in the vicinity of the amenities, and
        \item 
        the combination of transit amenities and high-density zoning results in a transient state of urbanization of poverty, followed by the formation of densely populated yet financially segregated urban core.
        This transition is delayed with increasing resident preference for community ties.
    \end{enumerate}
\end{enumerate}
All data and source code utilized in this work can be accessed online\footnote{at \href{https://github.com/jcmartinezmori/modeling_processes_of_neighborhood_change}{github.com/jcmartinezmori/modeling\_processes\_of\_neighborhood\_change}.}.

\subsection{Related Work}
\label{sec: related work}

Some models of neighborhood change phenomena are based on empirical statistics.
Chapple and Zuk~\cite{chapple2016forewarned} recounted experiences with the practical deployment of ``early warning systems'' for neighborhood change, which take the form of reports or online toolkits that monitor statistics of interest (e.g., tracking investments, population flows, housing sales).
However, they concluded that the current predictive power of these systems is generally poor, and that they are not well-integrated with the broader ``smart cities'' movement.
Rigolon and N\'emeth~\cite{rigolon2019toward} developed a regression model to identify predictors of gentrification for gentrification-susceptible neighborhoods. 
They consider variables roughly categorized into demographic, built environment, and policy related, and emphasize the importance of accounting for the contextual interplay across multiple variables.
Lindsey et al.~\cite{lindsey2021neighborhood} conduct a case study monitoring changes on gentrification indicators following the opening of new neighborhood amenities (in this case multiuse trails) in three different cities.
While they found some evidence of gentrification in all cases, they also found that not all gentrification-susceptible areas experienced gentrification; whether they did or not was localized and context-specific.
To summarize, statistical models may help identify risk factors for gentrification as a practical way of prioritizing policy efforts.
However, the complexity of these phenomena makes it challenging to extend this to a mechanistic understanding.

Another closely-related line of work is that of agent-based models of spatial segregation.
Schelling~\cite{schelling1971dynamic} developed agent-based models showing that agents with mild segregation preferences can nevertheless lead to highly segregated spatial patterns.
Grauwin et al.~\cite{grauwin2009competition} studied the effects of agent cooperation.
Fossett and Dietrich~\cite{fossett2009effects} studied the sensitivity of Schelling-like models of residential segregation on urban size, shape, and form.
More recently, Schelling-like models have been extended to more explicitly capture other concepts of neighborhood change.
Eckerd, Kim, and Campbell~\cite{eckerd2019gentrification} studied the influence of density and segregation preferences on spatial displacement.
Ortega, Rodr\'iguez-Laguna, and Korutcheva~\cite{ortega2021avalanches} studied the effects of sudden housing price hikes, namely their potential to kickstart an ``avalanche'' of spatial displacement.
Zhao and Randall~\cite{zhao2022heterogeneous} introduced a stochastic Schelling-like model to assess the effects of different spatial configurations of urban amenities on patterns of residential segregation.
Compared to traditional agent-based models, no-regret dynamics benefit from milder assumptions on agent behavior.
First, our domain-specific modeling is limited to the incentive structures perceived by residents, as opposed to hard-coded rules on their behavior or a priori assumptions on their state of equilibrium.
This represents a step forward from modeling agent behavior as if they were mechanical particles.
Moreover, as reviewed in Section~\ref{sec: technical background}, no-regret dynamics arise rapidly from a very natural process of learning by way of repeated play, which instills confidence on the predictive power of the resulting equilibrium concept.

\subsection{Discussion}
\label{sec: discussion}

\begin{figure}[ht]
     \centering
      \begin{subfigure}[b]{0.49\linewidth}
         \centering
         \includegraphics[trim={0cm 52.5cm 0 30cm},clip,width=\linewidth]{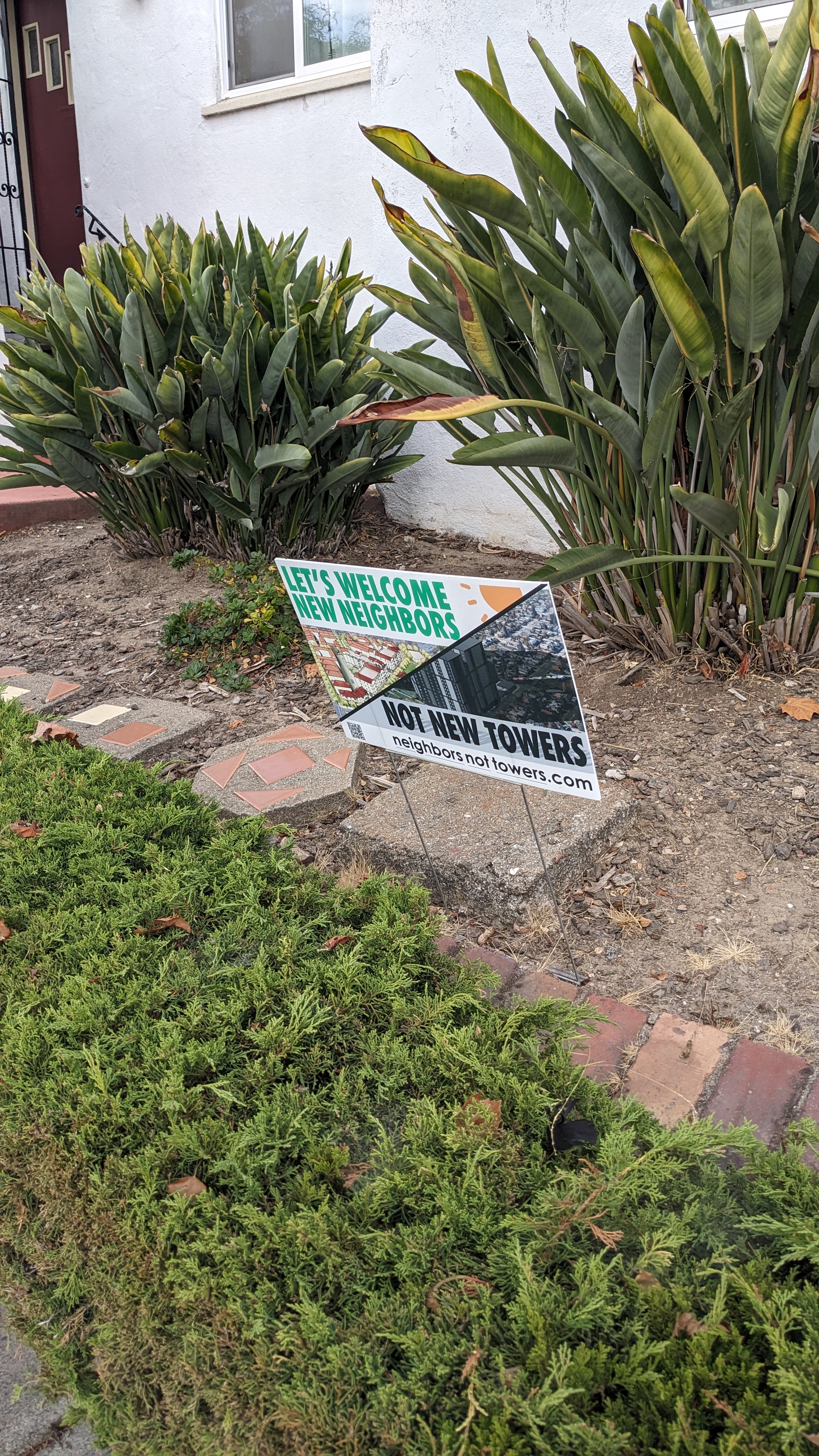}
         \caption{Coalitional bargaining}
         \label{fig: coalitional}
     \end{subfigure}
     \hfill
     \begin{subfigure}[b]{0.49\linewidth}
         \centering
         \includegraphics[trim={0 52.5cm 0 30cm},clip,width=\linewidth]{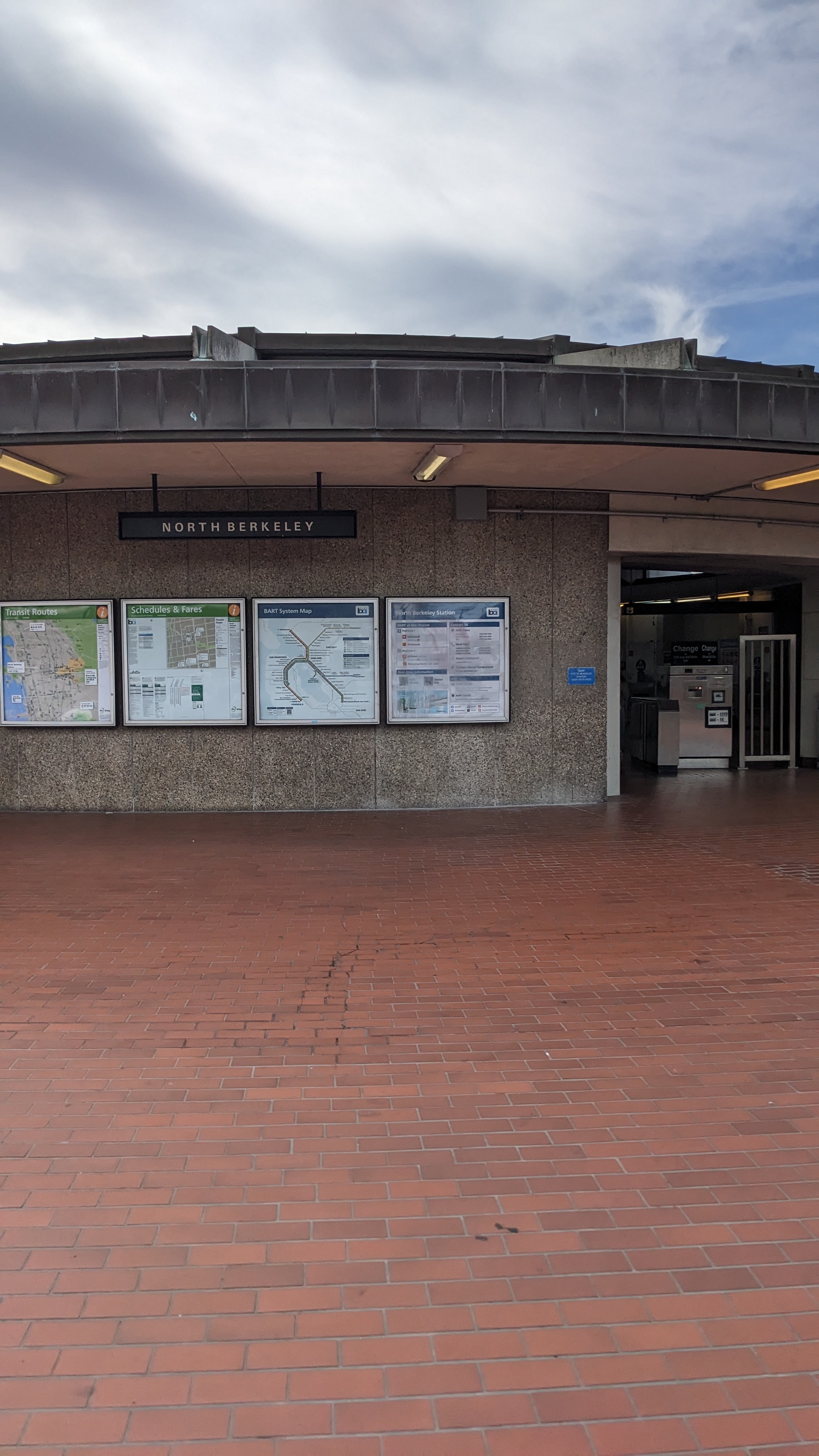}
         \caption{Transit infrastructure}
         \label{fig: transit}
     \end{subfigure}
     \caption{
     Organized opposition to high-density housing development in Berkeley, California.
     Figure~\ref{fig: coalitional}: A yard sign with the message ``Let's welcome new neighbors, not new towers'' and an accompanying website address. 
     Figure~\ref{fig: transit}: North Berkeley BART station serving the same neighborhood.
     \texttt{Source}: First author.
     }
     \label{fig: nimby}
\end{figure}

Our methodological approach and computational experiments offer stylized insights into the dynamics of transit-related neighborhood change.
However, it is inevitable that they have limitations.
We put these at the forefront of this work in an effort to transparentize our assumptions, contextualize our findings, mitigate unintended interpretations, and pinpoint areas requiring further investigation.
At a conceptual level, these include:
\begin{itemize}
    \item 
    No-regret dynamics assume full information repeated play (refer to Section~\ref{sec: technical background}).
    However, both full information and repeated play may be unrealistic.
    Partial information may be more realistic given resident lived experiences and access to online housing marketplaces, and from a technical point of view this relaxation would have little impact on our results.
    However, repeated play is much more difficult to justify, as housing decisions are made with much less frequency and taken much less lightly than the mathematical abstraction presumes.
    \item 
    The set of equilibria arising from no-regret dynamics is not necessarily unique; there may exist infinitely many.
    Therefore, at this stage, we can only argue conclusively about the equilibria obtained in our computational experiments.
    Typically, this kind of limitation is handled by characterizing the performance of worst-case equilibria (e.g., in the style of \emph{price-of-anarchy} bounds~\cite[Chapters 11 and 17]{roughgarden2016twenty}).
    However, doing so necessitates the adoption of a notion of ``system optimality.''
    In addition to the technical challenge, this poses a conceptual challenge in that there need not be an \emph{inherent} value judgment for processes of neighborhood change~\cite{lees2008gentrification,kohn2013wrong,zuk2018gentrification}.    
    \item 
    We neglect the potential for coalitional behavior. 
    Infrastructure and zoning policy is not done in a vacuum; their successful implementation is contingent on maintaining sufficient sociopolitical approval from groups of residents who may exhibit heterogeneous interests and heterogeneous bargaining power (e.g., as a result of home ownership status).
    Figure~\ref{fig: nimby} provides a case-in-point example of organized resistance to high-density housing development in a single-family home neighborhood that nevertheless enjoys large-scale transit infrastructure.
    Addressing this might require treatment from the perspective of cooperative games (refer to Myerson~\cite{myerson1991game}).
\end{itemize}

At a modeling level, we acknowledge that the drivers of housing decisions are much more nuanced than the aspects we account for (we further elaborate on these modeling assumptions in Section~\ref{sec: building a cost model} and Section~\ref{sec: case study}).
In addition, our experiments assume a closed system with no residential inflow or outflow (i.e., migration), and we isolate a particular urbanized region from a larger continuum of urbanized space, leading to artificial spatial aberrations (e.g., a site on the periphery of our study region may in fact neighbor more urbanized space).

Despite these limitations, we hope this work brings attention to the need for an improved understanding of\textemdash and eventual accounting for\textemdash the secondary effects of mathematics-enabled transportation planning.
Furthermore, we hope to have of made a case for the importance that these efforts cross disciplinary boundaries.

\subsection{Organization}
\label{sec: organization}

In Section~\ref{sec: technical background} we review the necessary technical background.
In Section~\ref{sec: a model of neighborhood change} we develop our model of neighborhood change.
In Section~\ref{sec: case study} we present our computational experiments.
We conclude in Section~\ref{sec: conclusion} with final remarks.

\section{Technical Background}
\label{sec: technical background}

The theory of \emph{no-regret dynamics} proposes a collective learning process through which a set of strategic agents rapidly reach a state of approximate equilibrium.
The background reviewed in this section is primarily based on Roughgarden~\cite[Chapters 13 and 17]{roughgarden2016twenty}.

The model begins by considering a single \emph{agent} in a game against an \emph{adversary}.
In this game, the agent has a fixed action space $A$ and enacts a sequence $a^1, a^2, \ldots, a^T \in A$ of actions over $T \in \mathbb{N}$ time steps.
At every time $t \in [T] \coloneqq \{1, 2, \ldots, T\}$, the agent adopts a \emph{mixed strategy} $p^t$ (i.e., a probability distribution) over $A$ and the adversary designs a cost function $c^t: A \rightarrow [0,1]$.
Once the agent enacts a random action $a^t$ according to $p^t$, they are revealed $c^t$ and incur cost $c^t(a^t)$.
The expected regret of a sequence $a^1, a^2, \ldots, a^T$ with respect to a fixed action $a \in A$ is
\begin{equation}
\label{eq: regret}
    \frac{1}{T}\left(\sum_{t=1}^T c^t(a^t) - \sum_{t=1}^T c^t(a)\right).
\end{equation}
An \emph{adaptive} adversary designs the cost function $c^t$ at time $t$ based only on previously enacted actions $a^1, a^2, \ldots, a^{t-1}$ and thus far observed mixed strategies $p^1, p^2, \ldots, p^{t-1}, p^t$.
How should the agent produce a sequence of actions against an adaptive adversary?

An online decision-making algorithm is said to be a \emph{no-regret algorithm} if, for every adaptive adversary and every fixed action $a$, the expected regret \eqref{eq: regret} given the sequence of actions it produces is $o(1)$ as $T \rightarrow \infty$.
Surprisingly, no-regret algorithms exist, and they are necessarily randomized.
Notably, the multiplicative weights update method (refer to Arora, Hazan, and Kale~\cite{arora2012multiplicative} for a survey on this method as a meta-algorithm) achieves expected regret $O(\sqrt{(\ln |A|)/T})$ with respect to every fixed action $a$.
Roughly, the method works by maintaining (initially uniform) weights over $A$, which are updated as time progresses.
At every time $t$, the algorithm normalizes these weights to obtain a mixed strategy $p^t$ over $A$, which it in turn uses to enact a random action $a^t$.
Then, it leverages the revealed cost function $c^t$ to discount the weight of (in hindsight) costly actions at a subtle but exponential rate\textemdash in this way achieving a suitable balance between exploration and exploitation.
We refer the reader to \cite[Chapter 17]{roughgarden2016twenty} for a formal description.
The same guiding principles extend to the \emph{bandit} model, where the agent is only revealed $c^t(a^t)$ as opposed to the entire cost function $c^t$, albeit with some additional sophistication~\cite[Chapters 6 and 7]{cesa2006prediction}.

No-regret dynamics arise in multi-agent settings in which agents independently and simultaneously employ their own instantiation of a no-regret algorithm.
The idea is that, while from the perspective of an individual agent the cost function $c^t$ revealed at time $t$ might as well be adversarial, in reality it results from the actions enacted by all other agents.

Formally, let $n \in \mathbb{N} \coloneqq \{1, 2, \ldots\}$ be the number of agents.
For the purposes of this work we assume the agents have an identical action space $A (= A_1 = A_2 = \cdots = A_n)$.
Let $\mathbf{a}^t = (a_1^t, a_2^t, \ldots, a_n^t) \in A^n$ encode the actions enacted by all agents at time $t$.
Similarly, let $\mathbf{a}_{-j}^t = (a_1^, \ldots, a_{j-1}^t, a_{j+1}^t, \ldots, a_n^t) \in A^{n-1}$ encode the actions enacted by all agents other than agent $j \in [n]$ at time $t$.
At time $t$, agent $j$ is revealed their \emph{individual} cost function $c_j^t: A \rightarrow [0, 1]$ given by
\begin{equation}
\label{eq: c_j^t}
    c_j^t(a) \coloneqq c_j(a, \mathbf{a}^t_{-j}).
\end{equation}
Note that \eqref{eq: c_j^t} is itself a function of $\mathbf{a}^t_{-j}$.
In this way, the cost function revealed to any given agent is coupled to the actions enacted by all other agents.

The main result of this theory is that no-regret dynamics rapidly converge to a state of approximate equilibrium.
To formally state this, let 
\begin{equation*}
    \sigma^t = \prod_{j=1}^n p_j^t
\end{equation*}
be the outcome distribution over agent actions at time $t$, where $p_j^t$ denotes the mixed strategy of agent $j$ at time $t$, and let
\begin{equation*}
    \sigma(T) = \frac{1}{T} \sum_{i=1}^T \sigma^t
\end{equation*}
be the time-averaged outcome distribution.
Then, $\sigma(T)$ is an approximate \emph{coarse correlated equilibrium} in the sense that, if after $T$ time steps each player achieves $\epsilon > 0$ regret, then
\begin{equation}
\label{eq: equilibrium}
    E_{\mathbf{a} \sim \sigma(T)}[c_j(\mathbf{a})] \leq E_{\mathbf{a} \sim \sigma(T)}[c_j(a', \mathbf{a}_{-j})] + \epsilon
\end{equation}
for each agent $j$ and each action $a'$.
In other words, no agent has a significant incentive to \emph{unilaterally} deviate from the time-averaged outcome distribution.

What remains is to determine, based on the application in mind, a form for the right-hand side of \eqref{eq: c_j^t} for each agent $j$.
That is, we need to design a cost function $c_j: A \times A^{n-1} \rightarrow [0,1]$ that is individual to agent $j$ and couples their perceived cost with the actions enacted by all other agents.
In this paper, we do so within the context of neighborhood change phenomena.

\section{A Model of Neighborhood Change}
\label{sec: a model of neighborhood change}

In this section we develop our model for the dynamics of neighborhood change.
As we formalize our model, we refer the reader to Table~\ref{tab: notation} for a summary of the notation used.

\begin{table}[ht]
\centering
\begin{tabular}{l|l}
Symbol & Description \\ \hline
$F$ & Set of (transit) amenity sites, indexed $f \in F$ \\
$H$ & Set of housing sites, indexed $h \in H$ \\
$\ell$ & Distance function $(F \cup U) \times (F \cup U) \rightarrow [0,1]$ \\
$R$ & Set of residents, indexed $j \in R$ \\
$T$ & Time horizon, indexed $t \in [T] \coloneqq \{1, 2, \ldots, T\}$ \\
$A_j^t (= H)$ & Action space of resident $j$ at time $t$ \\
$h_j^t$ & (Enacted) housing site of resident $j$ at time $t$ \\
$R^t(h)$ & $\{j \in R: h_j^t = h\}$ \\
$\mathcal{R}^t$ & $\{R^t(h)\}_{h \in H}$ \\
$\rho$ & Maximum number of residents allowed in a housing unit at a time \\
$0 < w_j < 1$ & Endowment of resident $j$ \\
$P^t(j,h)$ & Affordability ``score'' of housing site $h$ for resident $j$ at time $t$, ref.~\eqref{eq: P} \\
$L(h,F)$ & Amenity ``score'' of housing site $h$ with respect to $F$, ref.~\eqref{eq: L} \\
$W^t(j,h)$ & Community ``score'' of housing site $h$ for resident $j$ at time $t$, ref.~\eqref{eq: W} \\
$U^t(h)$ & Upkeep ``score'' of housing site $h$ at time $t$, ref.~\eqref{eq: U} \\
$0 \leq \lambda \leq 1$ & Relative importance of amenity and community ``scores,'' ref.~\eqref{eq: c} \\
$c_j^t$ & Cost function $A_j^t (= H) \rightarrow [0, 1]$ observed by resident $j$ at time $t$, ref.~\eqref{eq: c}
\end{tabular}
\caption{Summary of notation.}
\label{tab: notation}
\end{table}

\subsection{Preliminaries}
\label{sec: preliminaries}

Let $F$ and $H$ be finite sets of \emph{amenity} and \emph{housing} sites, respectively.
We treat $F$ and $H$ as points in a (possibly asymmetric) metric space equipped with a distance function $\ell: (F \cup U) \times (F \cup U) \rightarrow [0,1]$.
For example, $\ell$ may capture shortest-path distances in an underlying road network, normalized by its diameter (this is the case in Section~\ref{sec: case study}).

Let $R$ be a finite set of \emph{residents} and $T \in \mathbb{N}$ be a fixed time horizon.
We assume residents are free to move to any housing site of their choice at any point in time.
Formally, each resident $j \in R$ has the action space 
\begin{equation*}
    A_j^t = H
\end{equation*}
at every time $t \in [T]$.
Let $h_j^t \in H$ denote the \emph{enacted} housing site of resident $j$ at time $t$.
Similarly, for each housing site $h$, let $R^t(h) = \{j \in R: h_j^t = h\}$ denote the subset of residents housed at $h$ at time $t$.
Note that the collection $\mathcal{R}^t = \{R^t(h)\}_{h \in H}$ partitions $R$ for every $t$.

\subsection{Building a Cost Model}
\label{sec: building a cost model}

Based on the concepts of neighborhood change outlined in Section~\ref{sec: neighborhood change}, our goal is to design cost functions that capture the interactions between the following, as observed by residents:
\begin{enumerate}
    \item housing affordability,
    \item access to amenities\textemdash in particular transit amenities,
    \item community ties, and
    \item dwelling site abandonment/upkeep.
\end{enumerate}
We now describe our model to ``score'' resident actions with respect to each of these aspects.

\subsubsection{Housing Affordability}
To model affordability, we assume each housing unit is meant to house up to $\rho \in \mathbb{N}$ residents at a time, in this way reflecting any zoning ordinances on the maximum allowable residential density.
Moreover, we assume each resident $j$ has a fixed endowment $w_j \in (0,1)$.
We obtain $|R|$ distinct endowments as follows. 
We consider $|R|+1$ uniformly-spaced points in the $(0,1)$ interval, evaluate them along a \emph{Lorenz curve} $y: [0,1] \rightarrow [0,1]$ with the functional form
\begin{equation*}
    y = 1 - (1-x)^\frac{1}{2},
\end{equation*}
and take their consecutive differences.
A Lorenz curve summarizes the distribution of wealth in a society by encoding the cumulative fraction of total wealth as a function of the cumulative fraction of the population.
This particular functional form is a special case within the family of curves proposed by Raschel et al.~\cite{rasche1980functional} (namely, with parameters $\alpha=1/2$ and $\beta=1$).

We treat affordability as a binary attribute: housing site $h$ is affordable for resident $j$ at time $t$ if and only fewer than $\rho$ of the residents housed at $h$ at time $t$ have an endowment greater than $w_j$.
Formally, that is
\begin{equation}
\label{eq: P}
    P^t(j,h)
    =
    \begin{cases}
        1, & \text{if $|\{j' \in R^t(h): w_j' > w_j\}| < \rho$} \\
        0, & \text{otherwise.} 
    \end{cases}
\end{equation}

\subsubsection{Amenity Access}
We assume that, all else being equal, the closer a housing site $h$ is to an amenity site in $f \in F$, the more attractive $h$ is.
Formally, for each $h$ we define
\begin{equation}
\label{eq: L}
    L(h,F) = 1 - \min_{f \in F} \ell(h,f).
\end{equation}
Note that this quantity is independent of the spatial distribution of residents.
Moreover, it makes the qualitative assumption that amenities are \emph{relevant} to all agents.
Given our focus on transit amenities, this amounts to assuming residents are transit users.

\subsubsection{Community Ties}
We assume the endowment of residents housed near a housing site $h$ summarize the character of its community.
However, we note that this may or may not be a reasonable assumption.
This assumption may be justified on the basis of (commonly-held~\cite{kirkland2008what,fallon2021reproducing,rucks-ahidiana2022theorizing}) class-based conceptions of gentrification, which are characterized by the inflow of high-income households into low-income neighborhoods.
However, as argued by Sutton~\cite{sutton2020gentrification}, this ignores ``the fact that ethno-racial inequality operates alongside economic inequality.''
For example, while quantitative studies generally agree that newcomers tend to be high-income and/or highly-educated white households, the influence of and effects on original residents seem to be much more ethnoculturally nuanced ~\cite{mckinnish2010gentrifies,ellen2011low,sutton2020gentrification,rucks-ahidiana2021racial,rucks-ahidiana2022theorizing}.
In a similar vein, in summarizing previous studies on the movement middle-income black households into low-income black neighborhoods~\cite{boyd2005downside,pattillo2007black, moore2009gentrification}, Zuk et al.~\cite{zuk2018gentrification} posit that ``black in-movers feel more comfortable relocating to predominantly black neighborhoods because of a history of housing discrimination in predominantly white neighborhoods and the suburbs'' or do so as a purposeful form of ``racial uplift''~\cite{boyd2005downside}.
All in all, a general theory of neighborhood change cannot reduce the processes involved to purely class-based conceptions.

We nevertheless treat community as a purely class-based conception because: \textit{i}) it is simpler, both mathematically and otherwise; and \textit{ii}) it is a stepping stone toward improving our understanding of processes of spatial socioeconomic reorganization~\cite{rucks-ahidiana2022theorizing}.
Future work might improve this model alongside findings from social science research.

We assume that, all else being equal, the more similar the endowment of resident $j$ is to the endowment of residents near housing site $h$, the more attractive $h$ is to $j$.
We quantify the endowment of residents near $h$ at time $t$ as the weighted average
\begin{equation*}
    \frac{\sum_{j' \in R} w_j' (1 - \ell(h, h_{j'}^t))^2}{\sum_{j' \in R} (1 - \ell(h, h_{j'}^t))^2}.
\end{equation*}
In this way, the notion of ``nearness'' between $h$ and resident $j'$ at $t$ corresponds to the square of the proximity between $h$ and $h_{j'}^t$.
Accordingly, we define
\begin{equation}
\label{eq: W}
    W^t(j,h) = 1 - \left|w_j - \frac{\sum_{j' \in R} w_j' (1 - \ell(h, h_{j'}^t))^2}{\sum_{j' \in R} (1 - \ell(h, h_{j'}^t))^2}\right|.
\end{equation}

\subsubsection{Dwelling Abandonment/Upkeep}
Lastly, we equate the state of upkeep of a housing site $h$ to its state of occupancy or abandonment.
Formally, for each $h$ and each time $t$ we let
\begin{equation}
\label{eq: U}
    U^t(h)
    =
    \begin{cases}
        1, & \text{if $|R^t(h)| > 0$} \\
        0, & \text{otherwise.} 
    \end{cases}
\end{equation}
In this way, $h$ is in a state of upkeep at $t$ if and only if is not in a state of abandonment.

\subsection{Putting it Together}
\label{sec: putting it together}

We now combine Equations~\eqref{eq: P}-\eqref{eq: U}, the ``scores'' on resident actions with respect to each aspect, to obtain a single cost function as perceived (in hindsight) by residents.
We design this cost function so that it:
\begin{itemize}
    \item evaluates in the range $[0,1]$,
    \item evaluates to $1$ in the case of housing site unaffordability (as modeled in~\eqref{eq: P}) or state of abandonment (as modeled in~\eqref{eq: U}), which we treat as catastrophic forms of displacement, and
    \item ``smoothly'' transitions from $1$ to $0$ with decreasing amenity access (as modeled in~\eqref{eq: L}) or a decreasing sense of community (as modeled in~\eqref{eq: W}), which we treat as legitimate albeit more manageable forms of displacement.
\end{itemize}
In addition, we introduce a parameter $0 \leq \lambda \leq 1$ to reflect the relative importance residents attach to transit access (which increases with increasing $\lambda$) compared to a sense of community (which increases with decreasing $\lambda$).

For each resident $j$ and time $t$ we define $c_j^t: A_j^t(=H) \rightarrow [0,1]$ given by
\begin{equation}
\label{eq: c}
    c_j^t(h) = c_j(h, \mathbf{h}_{-j}^t) = 1 - P^t(j,h) \cdot U^t(h) \cdot e^{- \left(\lambda L(h,F) + (1 - \lambda)W^t(j,h)\right)}.
\end{equation}
Note that this model couples the perceived cost of a potential action of resident $j$ with the enacted actions $\mathbf{h}_{-j}^t$ of all other residents through the terms $P^t(j,h)$, $U^t(h)$, and $W^t(j,h)$. 
These in turn depend on the spatial distribution of residents at $t$, which is summarized by the collection $\mathcal{R}^t = \{R^t(h)\}_{h \in H}$.

\section{Case Study}
\label{sec: case study}

As mentioned in Section~\ref{sec: neighborhood change}, transit infrastructure is a notable class of large-capital intervention with the potential to drive neighborhood change. 
As such, it has been a subject of much attention in the planning literature; refer to Chapple and Loukaitou-Sideris~\cite{chapple2019transit} and the references therein.
In this section, we contribute to this literature with semi-quantitative insights on the spatial economics of transit-related neighborhood change.

We focus on ZIP code $11206$, which corresponds to part of the Williamsburg neighborhood of Brooklyn, New York.
We use the \texttt{networkx} package of Boeing~\cite{boeing2017osmnx} to retrieve a directed graph $(\mathcal{V}, \mathcal{A})$ representing its road network, with arcs weighted by length in meters.
Here, the edges correspond to road segments and the nodes correspond to their intersections.
We use these weights to obtain a normalized shortest-path distance function $\ell: \mathcal{V} \times \mathcal{V} \rightarrow [0,1]$.
Next, we retrieve the geographical coordinates of all Metropolitan Transportation Authority (MTA) subway stations in the area, map these stations to their nearest nodes in the graph, and mark these nodes as amenity sites $F \subseteq \mathcal{V}$.
We mark all other nodes $H = \mathcal{V} \setminus F$ as housing sites.
In total we obtain $|F| = 7$ and $|H| = 289$.

Before we further describe our computational experiments, we expand on our choice to focus on ZIP code $11206$.
For many years, Williamsburg has been a poster child of post-industrial gentrification in the United States~\cite{newman2006right,curran2007frying,martucci2024there}.
The drivers of this process are extremely complex, and we \emph{make no claim that our study gets even remotely close to capturing the full picture}.
In reality, we focus on this ZIP code because: \textit{i}) we need a subject region of moderate size that nevertheless exhibits an urbanized network structure, and \textit{ii}) we moreover need the region to exhibit a non-trivial configuration of transit amenities, in this case subway stations.
Therefore, while our stylized experiments shed some light on the underlying spatial economics, this should not be treated as the sole rationale behind real-world phenomena.

We conduct computational experiments with $|R| = |H| = 289$ residents and all possible combinations of parameters $\rho \in \{1, 2, 4, 8\}$ and $\lambda \in \{0.25, 0.75\}$.
In this way, we capture different levels of maximum allowable residential density as well as resident preferences over transit access and community ties.
For each choice of $\rho$ and $\lambda$, we initialize the experiment with the same randomness seed and execute no-regret dynamics using \eqref{eq: c}.
After $T$ time steps we obtain the product distribution $\sigma^{\rho, \lambda}(T)$ over $A^{|R|}$, which is an approximate coarse correlated equilibrium in the sense of \eqref{eq: equilibrium}.

\subsection{Experimental Results}
\label{sec: experimental results}

\begin{figure}[ht]
    \centering
    \begin{subfigure}[b]{0.2425\linewidth}
         \centering
         \includegraphics[width=\linewidth]{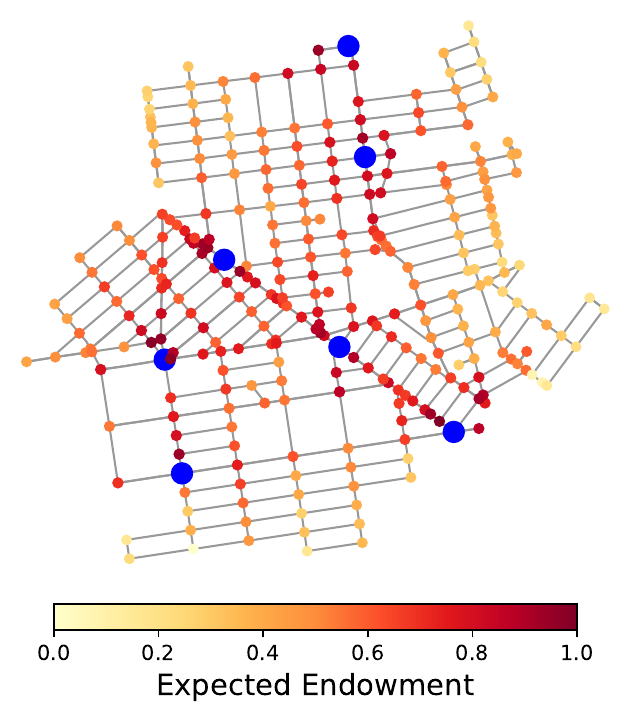}
         \caption{$\rho = 1$, $T=5000$}
         \label{fig: 0.25 1 5000}
    \end{subfigure}
    \hfill
    \begin{subfigure}[b]{0.2425\linewidth}
         \centering
         \includegraphics[width=\linewidth]{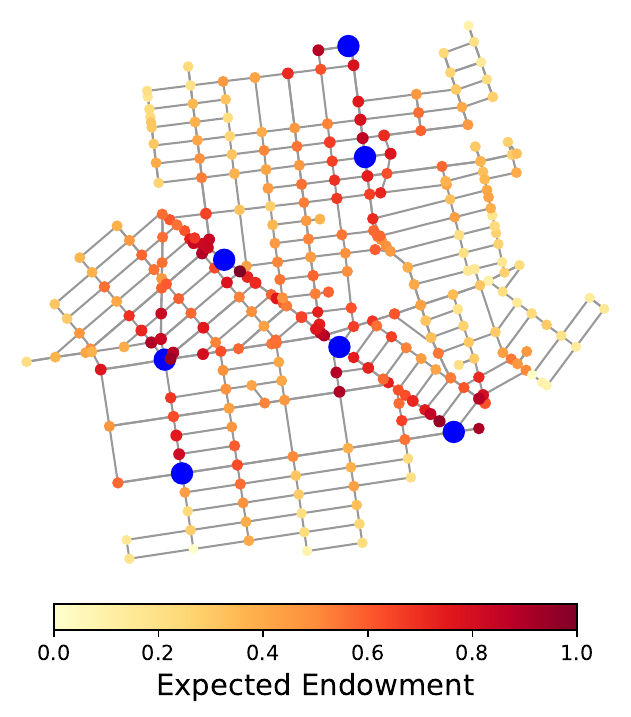}
         \caption{$\rho = 2$, $T=5000$}
         \label{fig: 0.25 2 5000}
    \end{subfigure}
    \hfill
    \begin{subfigure}[b]{0.2425\linewidth}
         \centering
         \includegraphics[width=\linewidth]{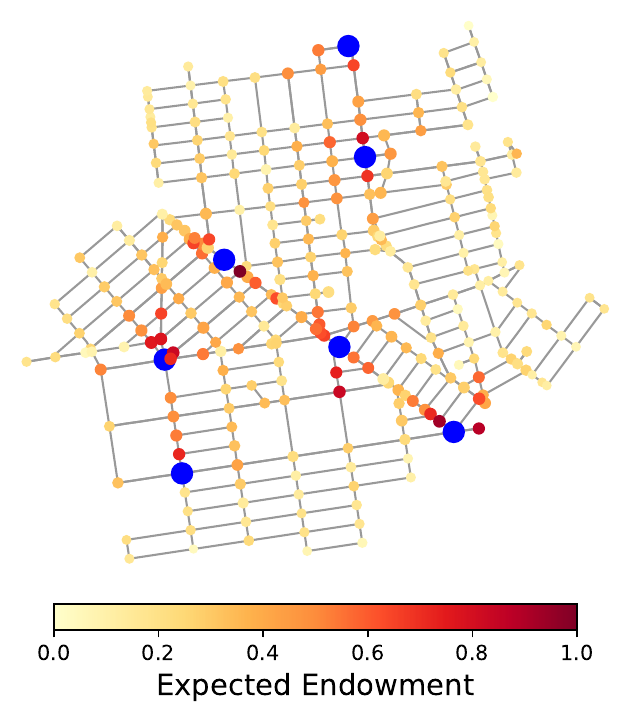}
         \caption{$\rho = 4$, $T=5000$}
         \label{fig: 0.25 4 5000}
    \end{subfigure}
    \hfill
    \begin{subfigure}[b]{0.2425\linewidth}
         \centering
         \includegraphics[width=\linewidth]{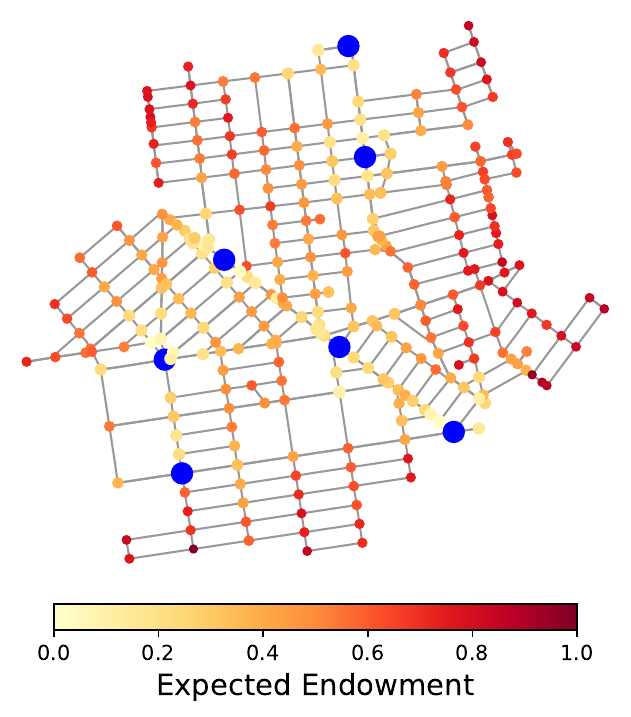}
         \caption{$\rho = 8$, $T=5000$}
         \label{fig: 0.25 8 5000}
    \end{subfigure}
    \hfill
        \begin{subfigure}[b]{0.2425\linewidth}
         \centering
         \includegraphics[width=\linewidth]{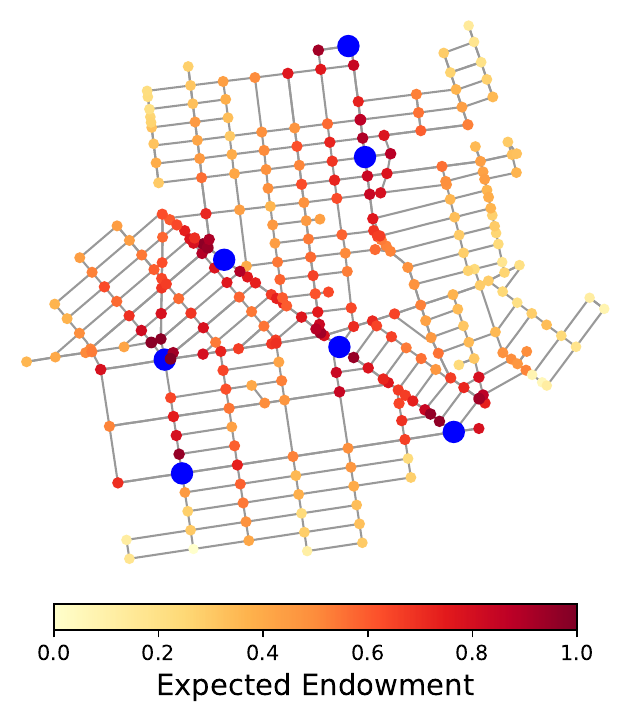}
         \caption{$\rho = 1$, $T=10,000$}
         \label{fig: 0.25 1 10000}
    \end{subfigure}
    \hfill
    \begin{subfigure}[b]{0.2425\linewidth}
         \centering
         \includegraphics[width=\linewidth]{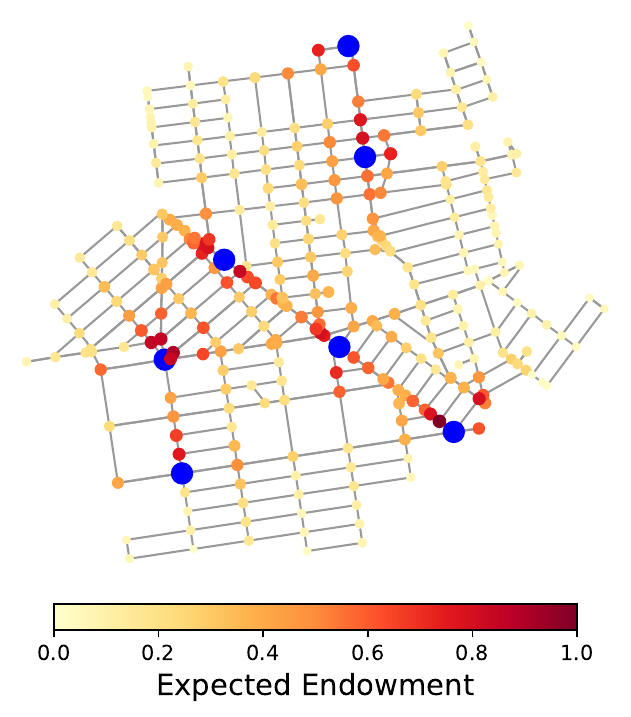}
         \caption{$\rho = 2$, $T=10000$}
    \end{subfigure}
    \hfill
    \begin{subfigure}[b]{0.2425\linewidth}
         \centering
         \includegraphics[width=\linewidth]{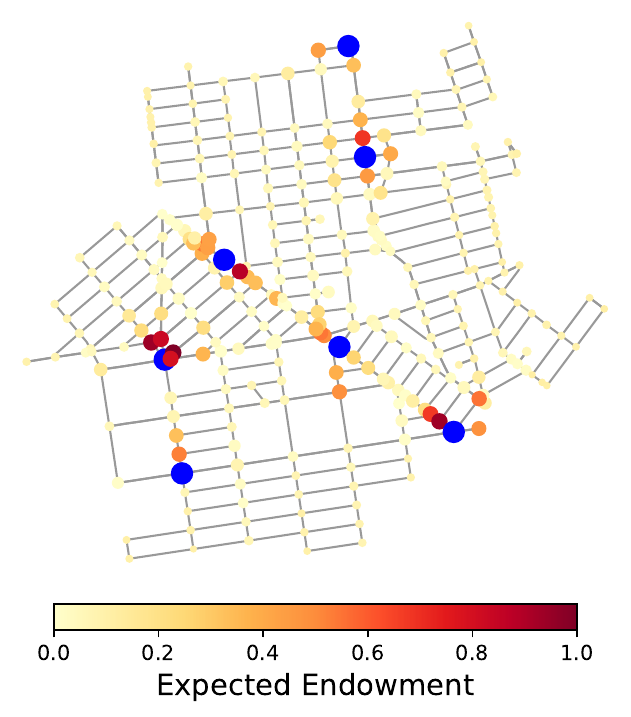}
         \caption{$\rho = 4$, $T=10000$}
    \end{subfigure}
    \hfill
    \begin{subfigure}[b]{0.2425\linewidth}
         \centering
         \includegraphics[width=\linewidth]{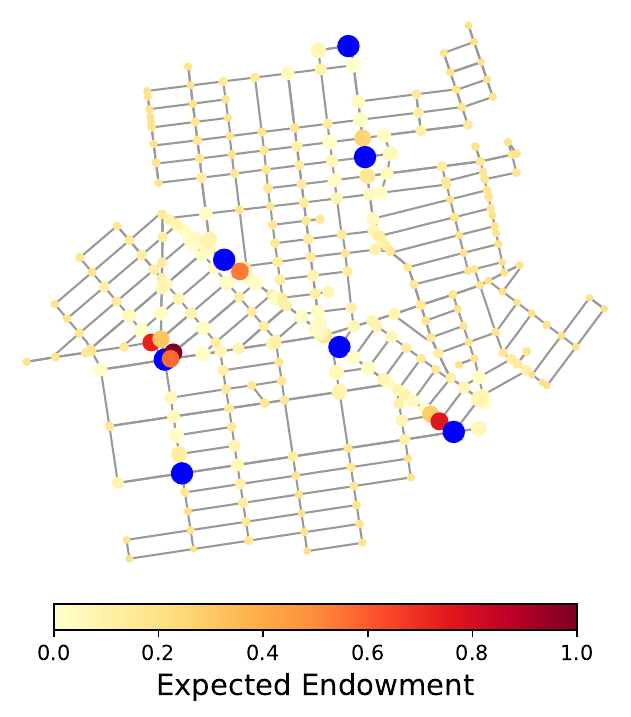}
         \caption{$\rho = 8$, $T=10000$}
         \label{fig: 0.25 8 10000}
    \end{subfigure}
    \hfill
    \begin{subfigure}[b]{0.2425\linewidth}
         \centering
         \includegraphics[width=\linewidth]{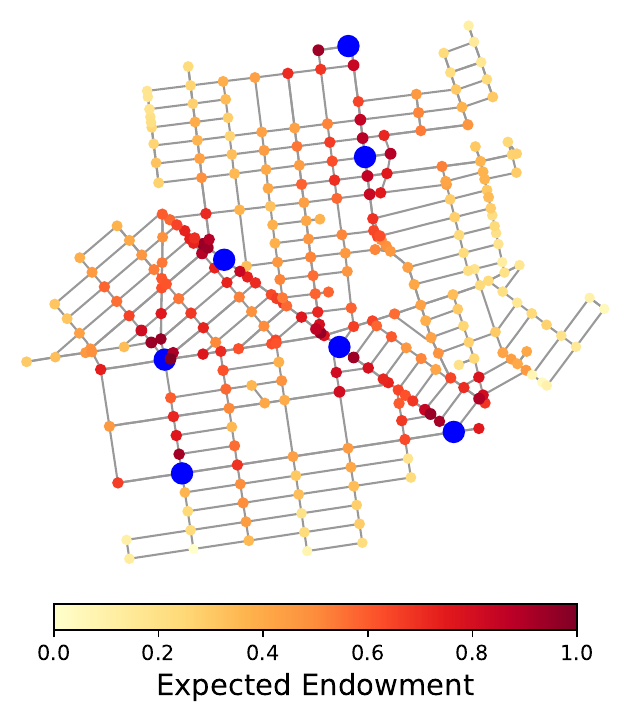}
         \caption{$\rho = 1$, $T=15000$}
         \label{fig: 0.25 1 15000}
    \end{subfigure}
    \hfill
    \begin{subfigure}[b]{0.2425\linewidth}
         \centering
         \includegraphics[width=\linewidth]{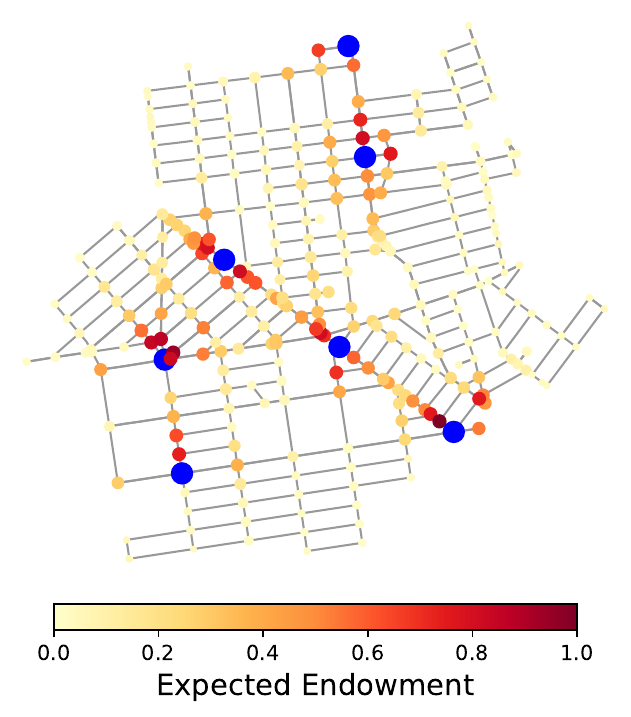}
         \caption{$\rho = 2$, $T=15000$}
         \label{fig: 0.25 2 15000}
    \end{subfigure}
    \hfill
    \begin{subfigure}[b]{0.2425\linewidth}
         \centering
         \includegraphics[width=\linewidth]{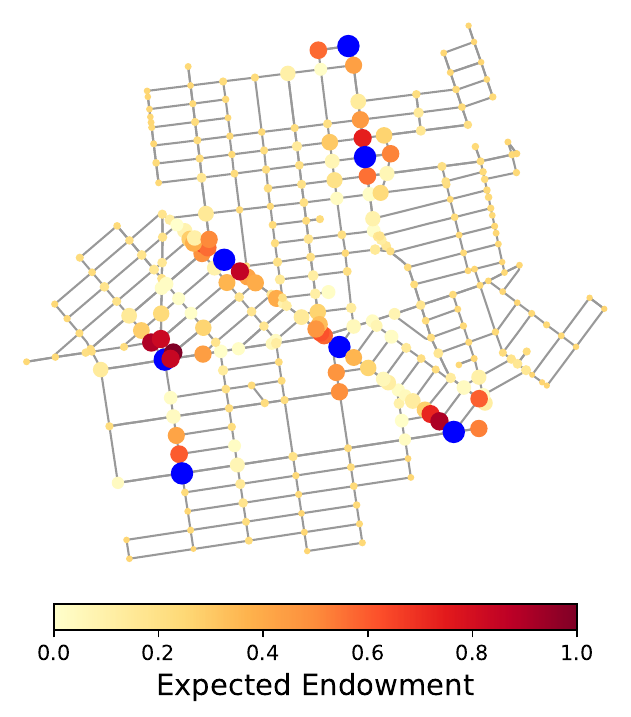}
         \caption{$\rho = 4$, $T=15000$}
         \label{fig: 0.25 4 15000}
    \end{subfigure}
    \hfill
    \begin{subfigure}[b]{0.2425\linewidth}
         \centering
         \includegraphics[width=\linewidth]{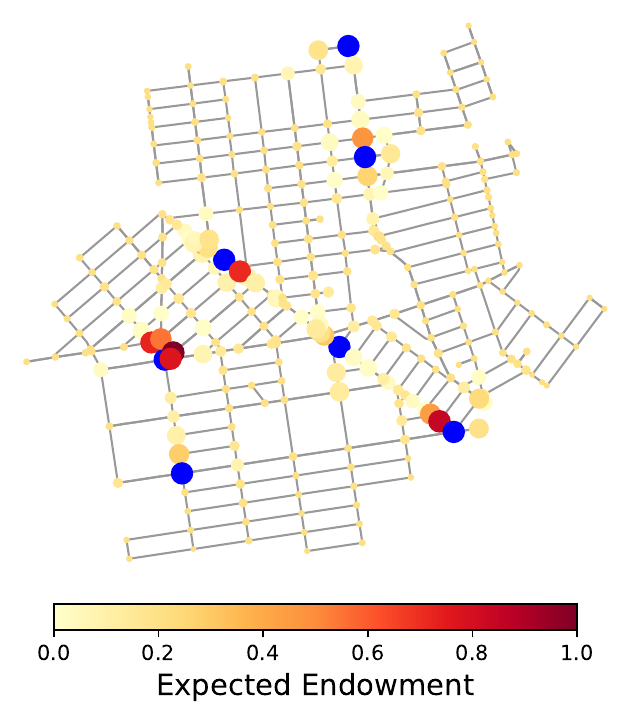}
         \caption{$\rho = 8$, $T=15000$}
         \label{fig: 0.25 8 15000}
    \end{subfigure}
    \caption{Equilibria for $\lambda = 0.25$ (i.e., residents favor amenity access)}
    \label{fig: lam 0.25}
\end{figure}

\begin{figure}[ht]
    \centering
    \begin{subfigure}[b]{0.2425\linewidth}
         \centering
         \includegraphics[width=\linewidth]{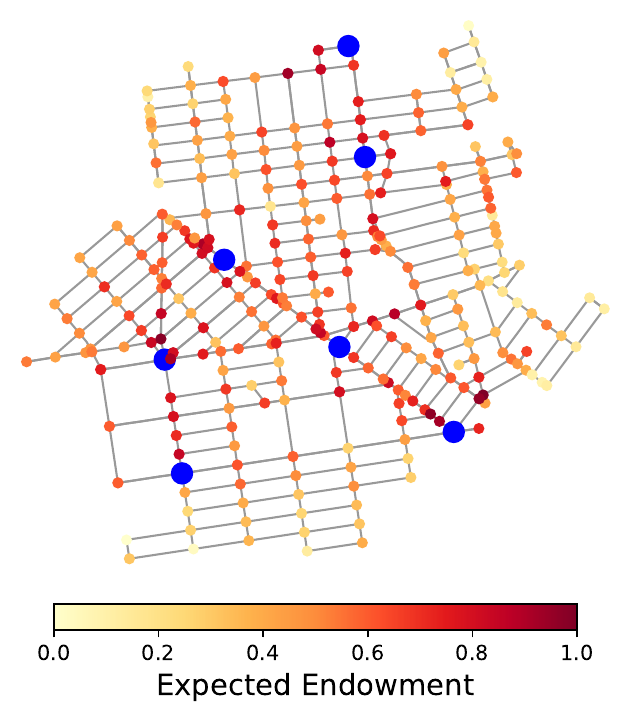}
         \caption{$\rho = 1$, $T=5000$}
         \label{fig: 0.75 1 5000}
    \end{subfigure}
    \hfill
    \begin{subfigure}[b]{0.2425\linewidth}
         \centering
         \includegraphics[width=\linewidth]{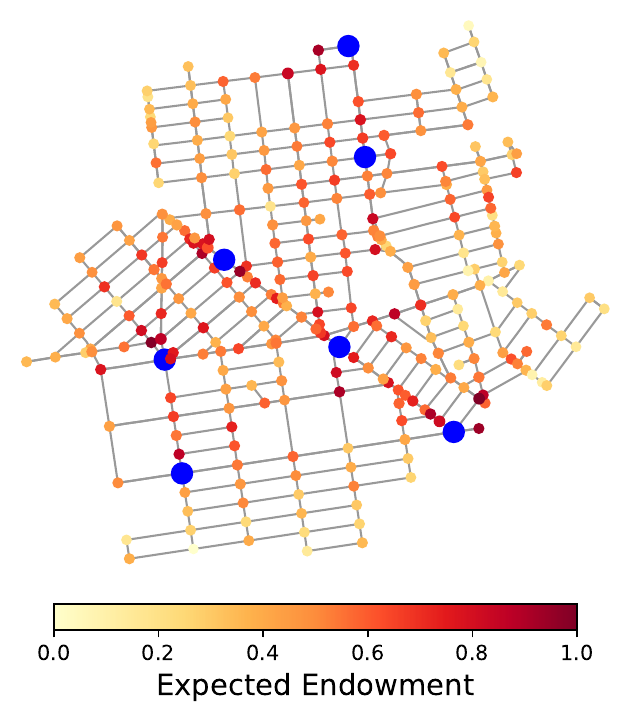}
         \caption{$\rho = 2$, $T=5000$}
         \label{fig: 0.75 2 5000}
    \end{subfigure}
    \hfill
    \begin{subfigure}[b]{0.2425\linewidth}
         \centering
         \includegraphics[width=\linewidth]{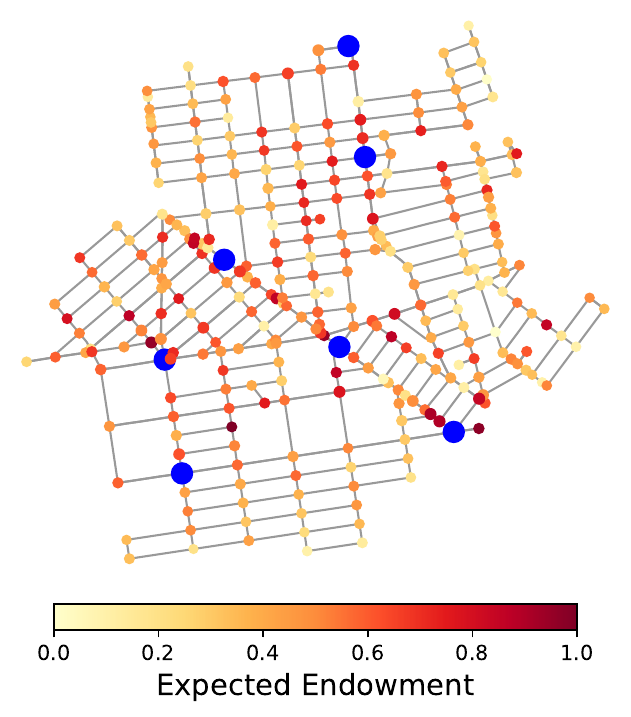}
         \caption{$\rho = 4$, $T=5000$}
         \label{fig: 0.75 4 5000}
    \end{subfigure}
    \hfill
    \begin{subfigure}[b]{0.2425\linewidth}
         \centering
         \includegraphics[width=\linewidth]{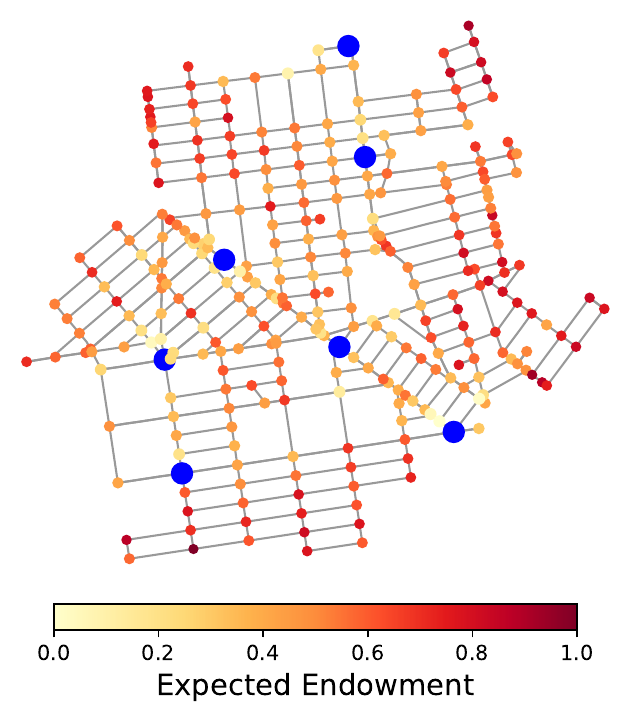}
         \caption{$\rho = 8$, $T=5000$}
         \label{fig: 0.75 8 5000}
    \end{subfigure}
    \hfill
        \begin{subfigure}[b]{0.2425\linewidth}
         \centering
         \includegraphics[width=\linewidth]{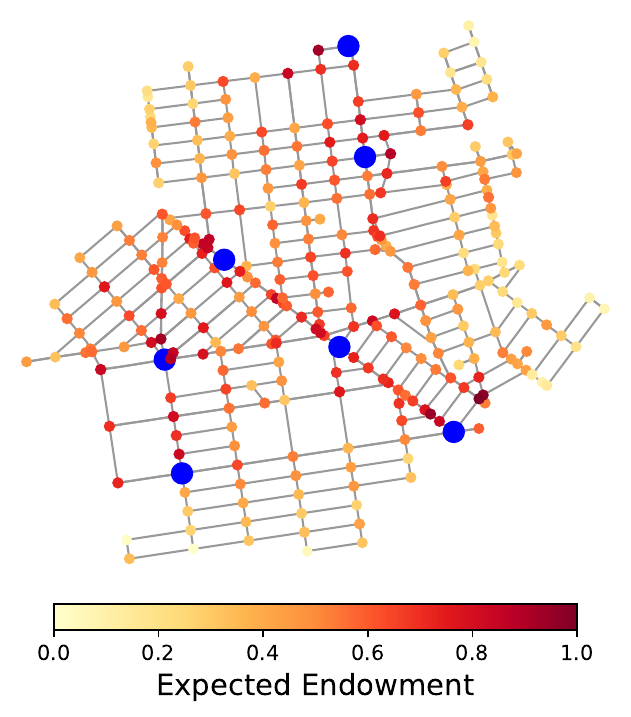}
         \caption{$\rho = 1$, $T=10000$}
         \label{fig: 0.75 1 10000}
    \end{subfigure}
    \hfill
    \begin{subfigure}[b]{0.2425\linewidth}
         \centering
         \includegraphics[width=\linewidth]{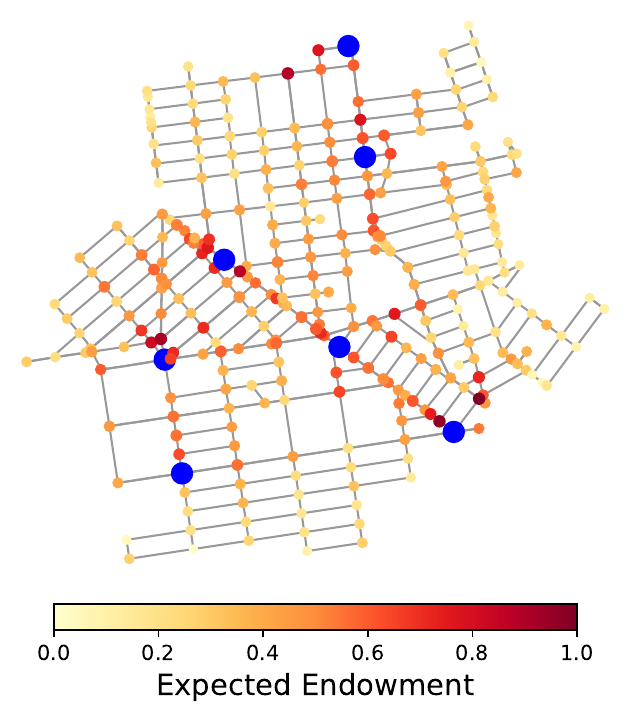}
         \caption{$\rho = 2$, $T=10000$}
    \end{subfigure}
    \hfill
    \begin{subfigure}[b]{0.2425\linewidth}
         \centering
         \includegraphics[width=\linewidth]{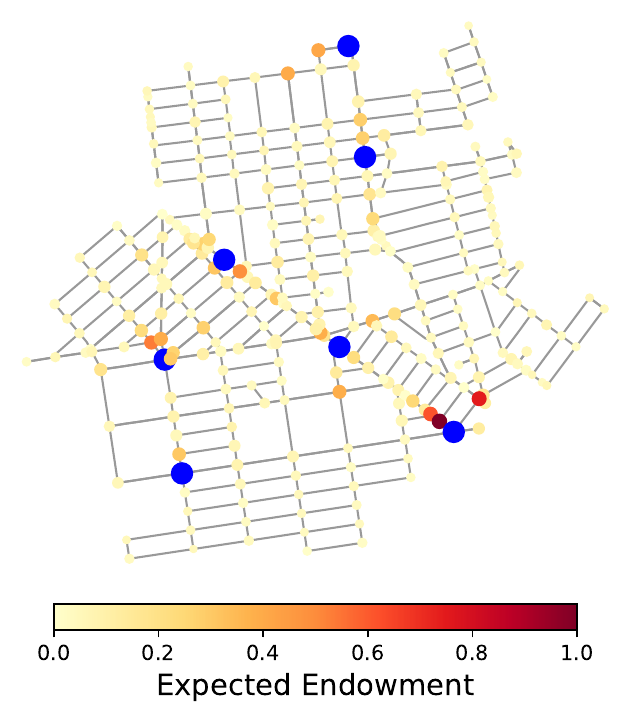}
         \caption{$\rho = 4$, $T=10000$}
    \end{subfigure}
    \hfill
    \begin{subfigure}[b]{0.2425\linewidth}
         \centering
         \includegraphics[width=\linewidth]{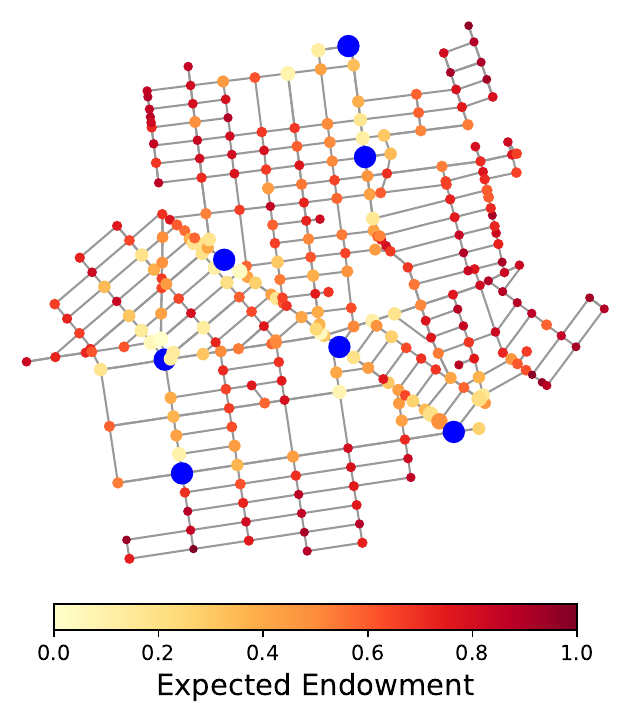}
         \caption{$\rho = 8$, $T=10000$}
         \label{fig: 0.75 8 10000}
    \end{subfigure}
    \hfill
    \begin{subfigure}[b]{0.2425\linewidth}
         \centering
         \includegraphics[width=\linewidth]{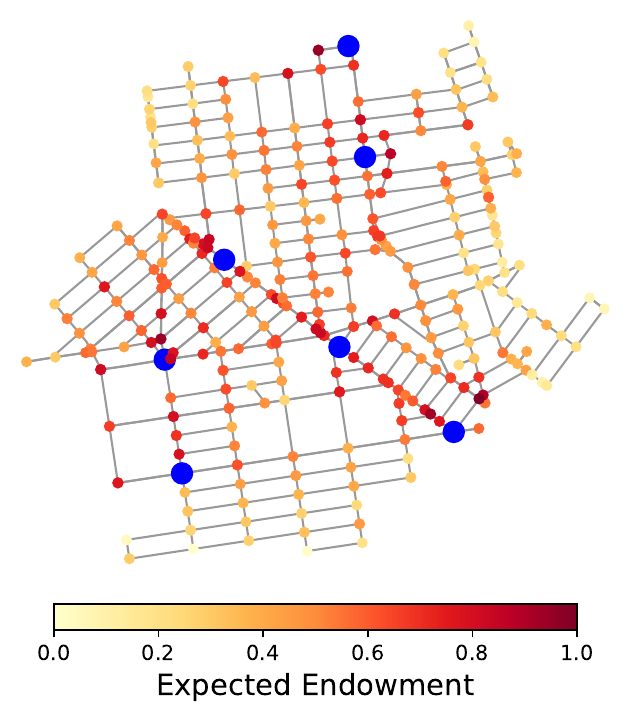}
         \caption{$\rho = 1$, $T=15000$}
         \label{fig: 0.75 1 15000}
    \end{subfigure}
    \hfill
    \begin{subfigure}[b]{0.2425\linewidth}
         \centering
         \includegraphics[width=\linewidth]{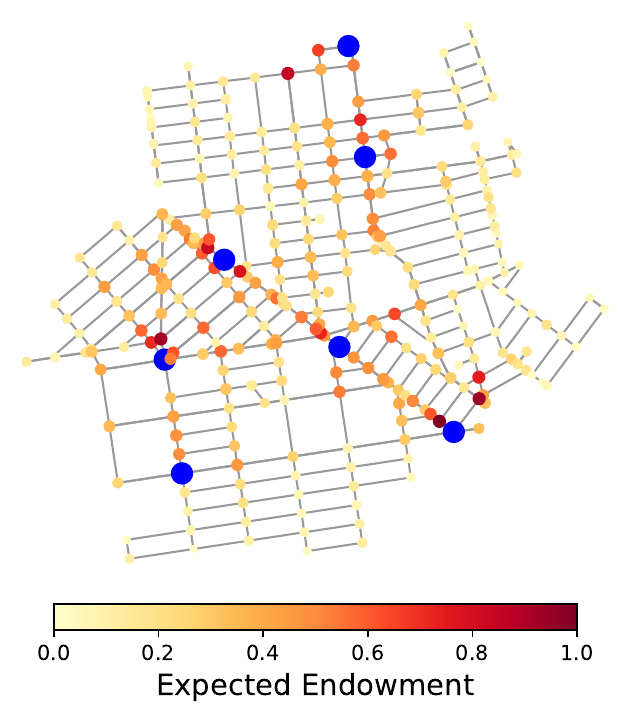}
         \caption{$\rho = 2$, $T=15000$}
         \label{fig: 0.75 2 15000}
    \end{subfigure}
    \hfill
    \begin{subfigure}[b]{0.2425\linewidth}
         \centering
         \includegraphics[width=\linewidth]{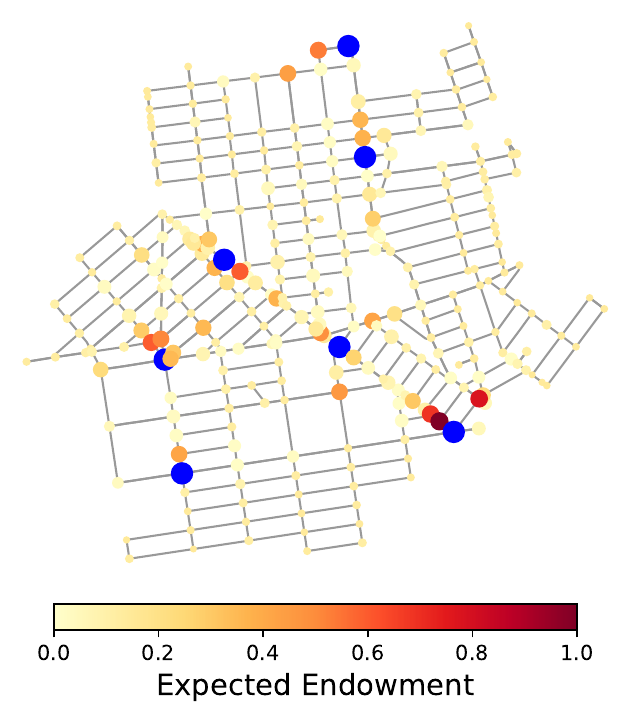}
         \caption{$\rho = 4$, $T=15000$}
         \label{fig: 0.75 4 15000}
    \end{subfigure}
    \hfill
    \begin{subfigure}[b]{0.2425\linewidth}
         \centering
         \includegraphics[width=\linewidth]{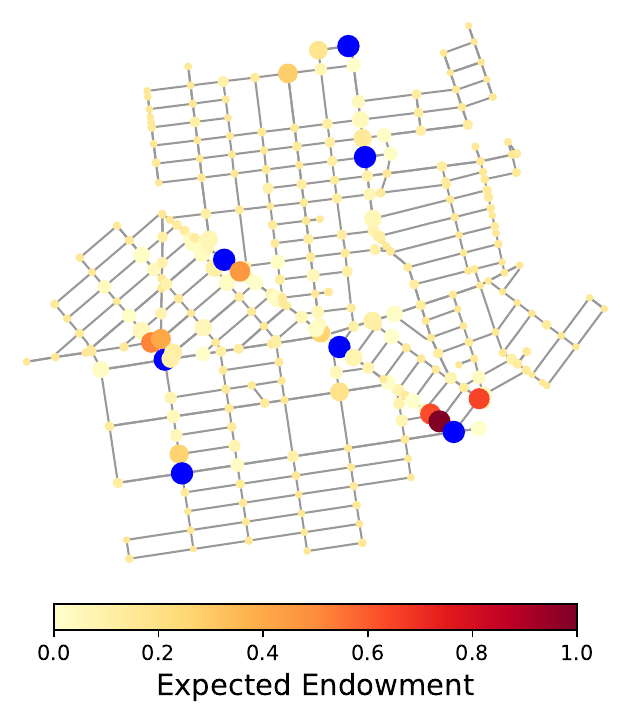}
         \caption{$\rho = 8$, $T=15000$}
         \label{fig: 0.75 8 15000}
    \end{subfigure}
    \caption{Equilibria for $\lambda = 0.75$ (i.e., residents favor community ties)}
    \label{fig: lam 0.75}
\end{figure}

In Figures~\ref{fig: lam 0.25} and \ref{fig: lam 0.75} we overlay snapshots of the resulting approximate equilibria $\sigma^{\rho, \lambda}(T)$ over the road network of ZIP code 11206, corresponding to $\lambda$ set to $0.25$ and $0.75$, respectively.
Within each figure, the rows correspond to increasing value of $\rho$ whereas the columns correspond to increasing value of $T \in \{5000, 10000, 15000\}$.
(We additionally considered the equilibria obtained for $T=20000$ and found them to be qualitatively similar to the equilibria obtained for $T=15000$).
The amenity sites $f \in F$ are depicted in solid blue while the housing sites $h \in H$ are depicted so that
\begin{itemize}
    \item their size is proportional to their expected population given $\sigma^{\rho, \lambda}(T)$, and
    \item 
    their color intensity, which ranges from yellow to orange to red, is normalized and proportional to their expected endowment given $\sigma^{\rho, \lambda}(T)$.
\end{itemize}

The figures consistently show that, with low-density zoning ($\rho = 1$), the closer a housing site is to a transit station, the greater its expected endowment (Figures~\ref{fig: 0.25 1 5000},~\ref{fig: 0.25 1 10000},~\ref{fig: 0.25 1 15000},~\ref{fig: 0.75 1 5000},~\ref{fig: 0.75 1 10000}, and~\ref{fig: 0.75 1 15000}).
However, the expected population of the housing sites remains fairly consistent over a wide geographical area surrounding the transit stations.
In other words, we observe that the \emph{combination of transit amenities and low-density zoning consistently results in the suburbanization of poverty}.

With medium-density zoning ($\rho = 2, 4$) we initially observe a geographically homogeneous expected population, combined with the suburbanization of poverty (Figures~\ref{fig: 0.25 2 5000},~\ref{fig: 0.25 4 5000},~\ref{fig: 0.75 2 5000}, and ~\ref{fig: 0.75 4 5000}).
However, over time, we observe a more pronounced concentration of residents in the vicinity of transit stations (Figures~\ref{fig: 0.25 2 15000},~\ref{fig: 0.25 4 15000},~\ref{fig: 0.75 2 15000}, and~\ref{fig: 0.75 4 15000}).
The expected endowment of housing sites near transit stations tends to become heterogeneous, as depicted by tones of orange in color intensity.
However, if residents favor community ties over amenity access ($\lambda = 0.75$), we observe some class-based segregation, especially for the higher end of the medium-density zoning spectrum ($\rho = 4$); compare the number of transit stations with a dark red vicinity in Figures~\ref{fig: 0.75 4 15000} and~\ref{fig: 0.25 4 15000}.
To summarize, we observe that the \emph{combination of transit amenities and medium-density zoning results in financially heterogeneous neighborhoods concentrated in the vicinity of the amenities. However, this heterogeneity decreases with increasing zoning density and increasing resident preference for community ties}.

Lastly, with high-density zoning ($\rho = 8$), our observations change depending on the choice of $T$ and $\lambda$.
In both figures, we initially observe a stage of \emph{urbanization} of poverty (Figures~\ref{fig: 0.25 8 5000},~\ref{fig: 0.75 8 5000}, and~\ref{fig: 0.75 8 10000}), followed by formation of a densely populated yet financially-segregated urban core (Figures~\ref{fig: 0.25 8 15000} and~\ref{fig: 0.75 8 15000}).
We observe that the transition between these stages is delayed with increasing resident preference for community ties ($\lambda=0.75$); compare Figure~\ref{fig: 0.75 8 10000} to Figure~\ref{fig: 0.25 8 10000}.
To summarize, we observe that the \emph{combination of transit amenities and high-density zoning results in a transient stage of urbanization of poverty, followed by the formation of densely populated yet financially segregated urban core.
This transition is delayed with increasing resident preference for community ties.}

\section{Conclusion}
\label{sec: conclusion}

In this paper, we proposed the use of game-theoretic concepts to study neighborhood change phenomena.
We formulated these processes as instances of no-regret dynamics, and mathematized concepts from the social sciences literature to model residential incentive structures around affordability, access to relevant transportation amenities, community ties (with respect to a class-based conception), and dwelling site abandonment/upkeep.
We showcased our model with computational experiments that shed some light on the spatial economics of transit-related neighborhood change; refer to Section~\ref{sec: contributions} for a summary of our findings.
Future work might address the limitations outlined in Section~\ref{sec: discussion}.
Ultimately, we hope methods in the spirit of this work might be developed as algorithmic tools for more equitable transportation infrastructure planning.

\section*{Acknowledgements}
Part of this research was performed while J.~C. Mart\'inez Mori and Zhanzhan Zhao were visiting the Mathematical Sciences Research Institute (MSRI), now becoming the Simons Laufer Mathematical Sciences Institute (SLMath), which is supported by NSF Grant No. DMS-1928930.
J.~C. Mart\'inez Mori is supported by Schmidt Science Fellows, in partnership with the Rhodes Trust.

\printbibliography

\end{document}